\documentclass[aps,prb,final,letterpaper,10pt,twocolumn,floats,showpacs,amsmath,amsfonts,amssymb]{revtex4-1}

\usepackage{graphicx}
\usepackage[colorlinks=true,citecolor=blue]{hyperref}

\begin{document}

\title{Local and non-local thermopowers in three-terminal nanostructures}

\author{G. Micha{\l}ek} \email{grzechal@ifmpan.poznan.pl}
\author{M. Urbaniak}
\author{B. R. Bu{\l}ka}
\affiliation{Institute of Molecular Physics, Polish Academy of Sciences, \\ ul. M. Smoluchowskiego 17, 60-179 Pozna\'{n}, Poland}

\author{T. Doma\'{n}ski}
\author{K. I. Wysoki\'{n}ski}
\affiliation{Institute of Physics, M. Curie-Sk{\l}odowska University, \\ pl. M. Curie-Sk{\l}odowskiej 1, 20-031 Lublin, Poland}

\date{\today}

\begin{abstract}
The thermoelectric effects in three-terminal structures with a quantum dot are considered. We propose the experimentally consistent protocol for determination of the transport coefficients in terms of the local and non-local conductances and thermopowers that can be measured in two steps, applying the `four probe technique'. This proposal is compared with other approaches discussed so far in the literature. As an example we study in detail the thermopower induced by the superconducting electrode in a subgap regime which might be useful for analysis of novel hybrid devices.
\end{abstract}


\maketitle

\section{Introduction}

Thorough description of the electronic transport in nanoscopic systems driven from their thermal equilibrium is important for designing such innovative thermodevices as: on-chip thermometers~\cite{feshchenko2015,giazotto2006}, heat to electricity converters~\cite{benenti2013}, sub-micrometer refrigerators~\cite{prance2009}, \textit{etc.} Thermoelectric properties have been so far examined mainly in two-terminal configurations~\cite{matveev2000, butcher1990, costi2010, dubi2011, krawiec2007, murphy2008, paulsson2003}, where the Seebeck coefficient is defined in a unique way. It probes a voltage needed to counterbalance a current induced by temperature difference developed across the system.

In the linear approximation, the current flowing between the left (\textit{L}) and right (\textit{R}) terminals is given by $J_{LR} = \mathcal{G} \Delta V_{LR} + \mathcal{G} S\Delta T_{LR}$, where $\mathcal{G}$ is an ohmic conductance, $\Delta V_{LR}$ is the voltage bias and $\Delta T_{LR}$ the temperature difference. Under the open circuit condition ($J_{LR} = 0$) the thermopower
\begin{equation} \label{def-tep}
S = - \left. \frac{\Delta V_{LR}}{\Delta T_{LR}} \right|_{J_{LR} = 0}
\end{equation}
measures a ratio between the voltage $\Delta V_{LR}$ induced by temperature difference $\Delta T_{LR}$. For the two-terminal junctions and bulk systems~\cite{mahan2000} this definition can be easily extended, even beyond the linear response regime~\cite{sanchez2013, lopez2013,svensson2013}. Seebeck coefficient (\ref{def-tep}) yields information, complementary to the electric conductance $\mathcal{G}$~\cite{segal2005}. In the simplest case, ${\mathcal{G}}$ is sensitive to the electron states at the Fermi energy ($E_F$), whereas the thermopower \textit{S} depends on a slope of the density of states near $E_F$, thus probing the particle-hole asymmetry~\cite{dubi2011}.

From an application point of view, the systems with good thermoelectric efficiency would be useful for waste energy harvesting~\cite{mahan1996,muhonen2012,sothmann2015}. As both the efficiency and the power output monotonically depend~\cite{benenti2013} on the thermoelectric figure of merit $ZT$ it is important to find bulk materials~\cite{snyder2008,zebarjadi2012} or heterostructures~\cite{heremans2008,duarte2009,trinh2008,machon2014}, where the large Seebeck effect guarantees high values of $ZT \propto S^2$.

In this regard very promising are multi-terminal nanostructures with an enhanced thermoelectric efficiency~\cite{mazza2014,wohlman2010,thierschmann2015,liu2010} especially under the broken time-reversal symmetry~\cite{wohlman2012,saito2011,zhang2013,brandner2013}. In these structures the non-equilibrium conditions are often accompanied by the important non-local effects requiring proper definitions of transport coefficients.

Generalization of the Ohm's law to multi-terminal systems has been pioneered by B\"{u}ttiker~\cite{buttiker1986} and resulted in an important distinction between local and non-local conductances $\mathcal{G}$ and resistances \textit{R} ($\mathcal{G} = R^{-1}$). He considered a ballistic transport between arbitrary leads \textit{via} the system coherently coupled to additional voltage probes. Non-local transport coefficients relate the currents between the selected terminals to voltage bias (or temperature difference) existing between the different electrodes. In particular, such non-local thermoelectric effects would be of interest for the energy harvesting devices~\cite{sothmann2015,thierschmann2015,szukiewicz2015}.

The definition (\ref{def-tep}) of the two-terminal thermopower, based on the condition $J_{LR} = 0$ is not directly applicable to the multi-terminal system due to other currents flowing in adjacent branches. Various three-terminal generalisations of (\ref{def-tep}) have been considered in the literature~\cite{mazza2014,machon2013,sanchez2011,wysokinski2012}. Although they all rely on the experimental feasibility the chosen conditions depend on the context.

\begin{figure} 
\includegraphics[width=\linewidth,clip]{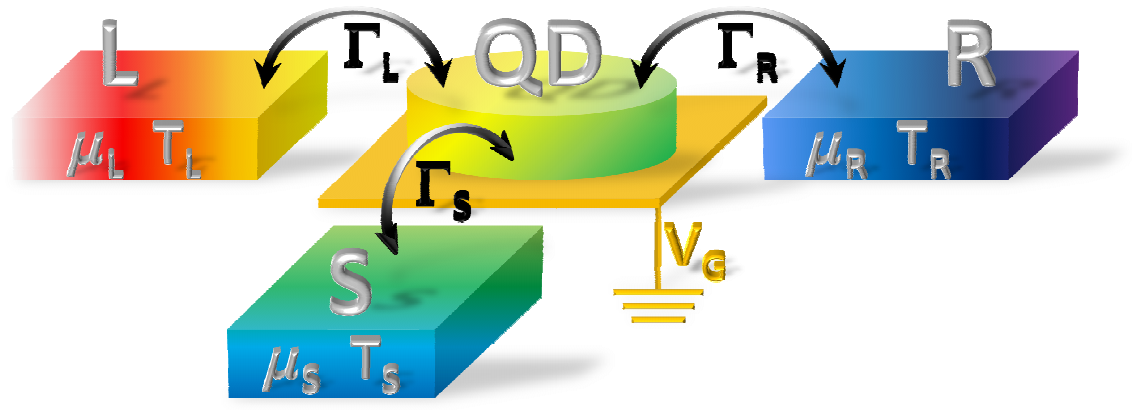}
\caption{\label{fig1}Schematic view of the nano-structure, consisting of the quantum dot (\textit{QD}) contacted to the left (\textit{L} and right \textit{R}) normal electrodes, characterized by temperatures $T_L$ and $T_R$ and chemical potentials $\mu_L$ and $\mu_R$. The third \textit{S} electrode, with temperature $T_S$ and the chemical potential $\mu_S$, can be either normal or superconducting. In the latter case \textit{S} is a source of the induced on-dot pairing.}
\end{figure}

The aim of our study is threefold. Firstly, we provide the experimentally consistent definitions of the local and non-local Seebeck coefficients valid for the three-terminal normal and hybrid devices by generalising B\"{u}ttiker~\cite{buttiker1986} approach to conductances. As a non-trivial application of our approach we analyse the thermopowers in the hybrid device sketched in Fig.~\ref{fig1}. Secondly, we study in detail the influence of the superconducting electrode on the subgap thermoelectric properties of the three-terminal devices. The superconducting electrode is responsible for strong non-local effects, previously observed in planar systems~\cite{russo2005,cadden2006,brauer2010,webb2012} and studied theoretically by various groups~\cite{futterer2009,schindele2014,machon2013,noh2013}. It may also cause the negative conductance, resulting from a competition between the crossed Andreev reflections and the direct electron tunneling involving normal electrodes~\cite{michalek2015}. This aspect is of an utmost importance for the efficient splitting of Cooper pairs~\cite{hofstetter2009,hermann2010,schindele2012}, spin filtering~\cite{braunecker2013}, or generation of the spin currents~\cite{he2014}. Finally, we confront our definition with other approaches discussed in the literature~\cite{machon2013,mazza2014,sanchez2011,wysokinski2012}.

The rest of the paper is organized as follows. In Sec.~\ref{sec:normal} we describe the formalism for determination of the charge currents and the corresponding transport coefficients: resistances/conductances and thermopowers. Based on such formalism we next generalize the thermopower Eq.~(\ref{def-tep}) for the three-terminal junctions, consistent with the experimentally measurable resistances. In Sec.~\ref{sec:hybrid} we study the local and non-local thermopowers of the hybrid three-terminal system with the quantum dot in presence of the superconducting lead. In Sec.~\ref{sec:other_definitions} we discuss other definitions of the thermopower known in the literature for multi-terminal systems. Summary and conclusions are given in Sec.~\ref{sec:summary} and the appendices provide technical details helpful for understanding the paper. In particular the theoretical model describing the system of Fig.~\ref{fig1} is presented in Appendix~\ref{app:model}.

\section{\label{sec:normal}Thermoelectric effects in three-terminal structures: general considerations}

Let us consider quantum dot (QD) coupled to three leads $i = \{ L, R, S \}$. Such nanostructure is the simplest realisation of the multi-terminal device with QD. We assume that the system is not-far from equilibrium, \textit{i.e.} with small temperature and chemical potential deviations ($\delta T_i$ and $\delta \mu_i$) with respect to reference values $\{ \mu, T \}$. As we are interested here in the conductances and thermopowers we shall consider only the charge currents $J_i$ flowing in aforementioned three terminals. They can be calculated using the Landauer-B\"{u}ttiker formalism~\cite{ness2010}. In the linear response regime $J_i$ can be expressed by the following general formula~\cite{jacquet2009}
\begin{equation} \label{eq:JLin2}
J_i = \sum_{j \neq i} \mathcal{L}_{ij, \mu} \Delta \mu_{ij} + \sum_{j \neq i} \mathcal{L}_{ij, T}
\Delta T_{ij} \; .
\end{equation}
where $eV_{ij} \equiv \Delta \mu_{ij} = \delta \mu_i - \delta \mu_j$, $\Delta T_{ij} = \delta T_i - \delta T_j$. In the absence of any magnetic field ($\mathbf{B} = \mathbf{0}$) and assuming the time reversal symmetry, the linear kinetic coefficients satisfy $\mathcal{L}_{ij, \mu} = \mathcal{L}_{ji, \mu}$ and $\mathcal{L}_{ij, T} = \mathcal{L}_{ji, T}$.

Any of the electrodes may be treated as the voltage probe \textit{P}, \textit{i.e.} the ideal voltmeter characterized by $J_P = 0$. Eq. (\ref{eq:JLin2}) establishes the relation between voltages, currents and thermal biases. It is used to relate the kinetic coefficients to the resistances and Seebeck coefficients measured for the considered device.

%
%

\subsection{\label{par:expcons}Protocol for measurements}

Here we shall discuss two basic measurements needed to get experimental information on both kinetic coefficients:  $\mathcal{L}_{ij, \mu} $ and $\mathcal{L}_{ij, T}$. To this end we assume that the system is coupled to some phonon bath~\cite{wohlman2010} and has well established and constant temperature. Using Eq. (\ref{eq:JLin2}) under the isothermal conditions, combined with the charge conservation $\sum_i J_i = 0$, one expresses first all potentials ($\Delta \mu_{ij}$) and later temperature differences ($\Delta T_{ij}$) in terms of the currents. As an example we consider the system with quantum dot and three electrodes shown in Fig.~\ref{fig1}.

\textit{First step} -- Under isothermal conditions $T_i = T$ one experimentally characterizes the charge transport, measuring the voltage $V_{ij}$ induced between terminals $\{ i, j \}$ in response to the current $J_{lk}$ applied between other terminals $\{l, k \}$~\cite{buttiker1986}. In this way, one can define 9 resistances $R_{lk, ij} \equiv V_{ij} / J_{lk}$ in three-terminal system, but only 3 of them are really independent. The resistances obey the symmetry relations $R_{lk, ij} = - R_{kl, ij} = - R_{lk, ji} = R_{kl, ji}$ and the reciprocity theorem which states that resistance measured in a four probe setup is invariant on the exchange of the voltage and current sources \textit{i.e.} $R_{lk, ij} = R_{ij, lk}$~\cite{buttiker1986} (see Ref. [\onlinecite{nazarov2009}] and Appendix~\ref{app:partialG} for details).

\textit{Second step} -- Having measured the resistances we relax the isothermal conditions and assume that (at least) two electrodes have different temperatures. Treating \textit{S} electrode as the voltage probe ($J_S = 0$) the charge conservation implies $J_L = - J_R \equiv J_{LR}$. The bias between $m$ and $n$ electrodes is calculated from \eqref{eq:JLin2} and reads
\begin{align}
\Delta \mu_{mn} & = e R_{mn, LR} J_{LR} + s_{mn, LR} \Delta T_{LR} \nonumber \\
& + s_{mn, RS} \Delta T_{RS} + s_{mn, SL} \Delta T_{SL} \; ,
\label{eq:mu1212s}
\end{align}
where $s_{mn, ij} \equiv - e R_{mn, ij} \mathcal{L}_{ij, T}$. Let us stress that the resistances $R_{mn, ij}$ have been determined in the first step of the measurement procedure. They are needed for determinations of the local and non-local thermopowers. The equations analogous to (\ref{eq:mu1212s}) can be derived for \textit{L} or \textit{R} beeing the floating electrodes. In general, one has 9 equations with 3 independent parameters $\mathcal{L}_{ij, T}$, or equivalently with 3 different coefficients $s_{mn, ij}$. Note that only two temperature gradients are really independent, because $\Delta T_{LR} = \Delta T_{LS} - \Delta T_{SR}$.

Taking into account the relation \eqref{eq:mu1212s} we see that the measurement of the voltage $\Delta \mu_{mn}$ induced by the temperature difference $\Delta T_{LR}$ gives the Seebeck coefficient. Let us recall that standard definition of the Seebeck coefficients %
\begin{equation}
S_{mn, ij} = - \left. \frac{\Delta \mu_{mn}}{e \Delta T_{ij}} \right|_{\{ cond \}}
\label{def:theropower}
\end{equation}
requires appropriate experimental conditions $\{ cond \}$. As already mentioned in the Introduction the only condition in the two-terminal system is the vanishing of the current. In multi-terminal devices there exist a large number of various possibilities. The only possible consistent set of conditions may be read off from the equation (\ref{eq:mu1212s}). One can assume that all temperature differences and the current take on nonzero values, or only one of them \textit{e.g.} $\Delta T_{LR}$ is non-vanishing. In the latter case the current may be required to vanish or can also be measured. Depending on the assumed condition one measures slightly different value of the the local and non-local Seebeck coefficients.

The latter procedure for the measurement of the Seebeck coefficients (\ref{def:theropower}) is consistent with the four probe method of measuring the resistances in multi-terminal systems~\cite{buttiker1986,nazarov2009}. This important consistency of our approach differs from other generalisations for local and non-local thermopowers considered in the papers~[\onlinecite{mazza2014,machon2013,sanchez2011,wysokinski2012}].

\subsection{Determination of the thermal linear coefficients}

The proposed scheme for measuring the resistances and thermopowers allows for experimental determination of the linear coefficients $\mathcal{L}_{LR, T}, \mathcal{L}_{LS, T}$ and $\mathcal{L}_{RS, T}$ and consistent comparison with theoretical models. For the considered setup at least 3 additional measurements are therefore required, which in the following we propose to perform under open circuit conditions (imposing $J_i = 0$ for all electrodes). Note, that this assumption makes our condition in (\ref{def:theropower}) similar to that of by Mazza \textit{et al.}~\cite{mazza2014}.

There exist two possible set-ups in our system. In the first case one takes \textit{e.g.} the condition $\Delta T_{SL} = 0$, while in the second $\Delta T_{RS} = 0$. Then one measures the potential differences: $\Delta \mu_{LR}$, $\Delta \mu_{RS}$ and $\Delta \mu_{SL}$ and the Seebeck coefficients $S_{mn, ij}$ for the temperature differences $\Delta T_{RS} = - \Delta T_{LR}$, $\Delta T_{SL} = - \Delta T_{LR}$, respectively. Solving the system of equations \eqref{eq:mu1212s} for $s_{mn, ij}$ one obtains
\begin{align} \label{eq:slrlr}
s_{LR, LR} & = \frac{e R_{LR, LR}}{D_R} ( R_{LR, RS} S_{RS, LR} - R_{RS, RS} S_{LR,LR} ) \; , \\
s_{SL, SL} & = \frac{e R_{LR, SL}}{D_R} ( R_{LR, RS} S_{RS, LR} - R_{RS, RS} S_{LR, LR} ) \nonumber \\ \label{eq:sslsl}
& + e S_{SL, LR} \; , \\ \label{eq:srsrs}
s_{RS, RS} & = \frac{e R_{RS, RS}}{D_R} ( R_{LR, LR} S_{RS, LR} - R_{LR, RS} S_{LR, LR} ) \; ,
\end{align}
where $R_{LR, LR} = - R_{LR, RS} - R_{LR, SL}$, $R_{RS, RS} = - R_{RS, LR} - R_{RS, SL}$ [see Appendix~\ref{app:partialG}]. The asymmetry between Eqs.\ (\ref{eq:sslsl}) and (\ref{eq:srsrs}) comes from different conditions $\Delta T_{SL} = 0$ or $\Delta T_{RS} = 0$ for which $\Delta \mu_{LR}$, $\Delta \mu_{RS}$ or $\Delta \mu_{SL}$ have to be measured. Consequently one can determine $\mathcal{L}_{ij, T}$ from experimental measurements of the Seebeck coefficients and resistances using the relation
\begin{equation}
\mathcal{L}_{ij, T} = - \frac{s_{mn, ij}}{e R_{mn, ij}} \; .
\end{equation}
This algorithm can be adopted to any kind of electrodes (magnetic, superconducting, \textit{etc.}) and any number of terminals. In the next section we shall illustrate how it captures the strong non-local thermoelectric effects driven by the anomalous Andreev scattering in the three-terminal structures comprising the superconducting reservoir interconnected, through the quantum dot, to two metallic electrodes.

\section{\label{sec:hybrid}Example: hybrid structures with superconducting lead}

Let us consider the local and non-local thermopowers of three-terminal hybrid system with the quantum dot placed between two metallic electrodes (\textit{L} and \textit{R}) and the superconducting reservoir (\textit{S}) as shown in Fig.~\ref{fig1}. We apply the formalism introduced in the preceding section, focusing on the linear response regime when the (subgap) transport is strongly affected by the Andreev scattering processes. The important problem of the gauge invariance of the theory is solved~\cite{michalek2015} by measuring chemical potentials of the normal electrodes and on dot energy level from the chemical potential $\mu_S$ of the superconducting electrode. It will be set to zero, unless specified otherwise. It has to be stressed that in this section of special interest are various processes in the system and their contributions to both sets of kinetic coefficients.

For voltages much smaller than the superconducting energy gap $\Delta$ the subgap current $J_L$ consists of the following three contributions~\cite{michalek2013}
\begin{align} \label{currET}
J_L & = \frac{2e}{h} \int dE \, T^{ET}(E) \, [ f_L(E) - f_R(E) ] \nonumber \\
& + \frac{2e}{h} \int dE \, T^{DAR}(E) \, [ f_L(E) - \tilde{f}_L(E) ] \nonumber \\
& + \frac{2e}{h} \int dE \, T^{CAR}(E) \, [ f_L(E) - \tilde{f}_R(E) ] \;,
\end{align}
where $f_\alpha (E) = \{ \exp [ (E - \mu_\alpha) / k_B T_\alpha ] + 1 \}^{-1}$ and $\tilde{f}_\alpha (E) = 1 - f_\alpha (-E) = \{ \exp [ (E + \mu_\alpha) / k_B T_\alpha ] + 1 \}^{-1}$ denote the Fermi-Dirac distribution functions for electrons and holes, respectively. The first part describes usual electron tunneling (ET) between \textit{L} and \textit{R} electrodes with the transmittance $T^{ET} (E) = \Gamma_L \Gamma_R |G_{11}^r (E)|^2$, whereas the second and the third parts characterize direct (DAR) and crossed (CAR) Andreev reflection processes with the corresponding transmittances $T^{DAR} (E) = \Gamma_L^2 | G_{12}^r (E) |^2$ and $T^{CAR} (E) = \Gamma_L \Gamma_R | G_{12}^r (E) |^2$. These functions $T^\kappa (E)$ depend on the couplings $\Gamma_i$ and on the (diagonal or off-diagonal) elements of the QD Green's function $\hat{G}^{r} (E)$ in the Nambu representation (see Appendix~\ref{app:model} and Ref.~[\onlinecite{michalek2013}] for details). The current $J_R$ is expressed by the formula analogous to \eqref{currET} by exchanging the indices $L \leftrightarrow R$.

For small perturbations $\delta \mu_\alpha \equiv \mu_\alpha - \mu_S$ and $\delta T_\alpha \equiv T_\alpha - T_S$ we can expand the current $J_L$ as
\begin{align} \label{eq:jsc}
J_L & = \mathcal{L}_{LR, \mu}^{ET} \left( \delta \mu_L - \delta \mu_R \right) + \mathcal{L}_{LR, \mu}^{CAR} \left( \delta \mu_L + \delta \mu_R \right) \\
& + 2 \mathcal{L}_{LL, \mu}^{DAR} \delta \mu_L + \left( \mathcal{L}_{LR, T}^{ET} + \mathcal{L}_{LR, T}^{CAR} \right) \left( \delta T_L - \delta T_R \right) \nonumber
\end{align}
with the coefficients $\mathcal{L}_{\alpha \beta, \gamma}^\kappa$ referring to the process $\kappa = \{ ET, DAR, CAR \}$; the subscripts correspond to $\alpha, \beta = \{ L, R \}$ and $\gamma = \{ \mu, T \}$, respectively. The linear coefficients $\mathcal{L}_{\alpha \beta, \gamma}^\kappa$ can be obtained from (\ref{currET}) and they read
\begin{align} \label{Lmucal}
\mathcal{L}_{\alpha \beta, \mu}^\kappa & = \frac{2e}{h} \int dE \, T^\kappa (E) \left[ - \frac{\partial f}{\partial E} \right] \; , \\ \label{LTcal}
\mathcal{L}_{\alpha \beta, T}^\kappa & = \frac{2e}{hT} \int dE \, E \, T^\kappa (E) \left[ - \frac{\partial f}{\partial E} \right] \; .
\end{align}
The coefficient $\mathcal{L}_{LR, \mu}^{ET}$ is related to the voltage induced by processes transferring single electron between the metallic \textit{L} and \textit{R} leads. We call this process the electron transfer (ET). The next term $\mathcal{L}_{LL, \mu}^{DAR}$ corresponds to the direct Andreev reflection, when electron from the normal \textit{L} lead is converted into the Cooper pair (in the \textit{S} electrode) and the hole is reflected back to the same lead \textit{L}. The coefficient $\mathcal{L}_{LR, \mu}^{CAR}$ corresponds to the non-local crossed Andreev reflection, when a hole is reflected to the second \textit{R} lead. The other set of linear coefficients $\mathcal{L}_{ij, T}^\kappa$ provides thermal contributions to the current by the process $\kappa = ET, CAR$. Note, that $\mathcal{L}_{LL, T}^{DAR}$ is absent in \eqref{eq:jsc} because an incident electron and a reflected hole stem from the same electrode.

\subsection{Local and non-local thermopowers}

Once the local and non-local resistances of the hybrid system are determined under isothermal conditions (see Appendix~\ref{app:res}) we can express the local and non-local thermopowers as
\begin{align} \label{eq:L-float-LS}
S_{LS} & \equiv - \left. \frac{\Delta \mu_{LS}}{e \Delta T_{RL}} \right|_0 = R_{LS, RL} ( \mathcal{L}_{LR, T}^{ET} + \mathcal{L}_{LR, T}^{CAR} ) \; , \\ \label{eq:L-float-RS}
S_{RS} & \equiv - \left. \frac{\Delta \mu_{RS}}{e \Delta T_{RL}} \right|_0 = R_{RS, RL} ( \mathcal{L}_{LR, T}^{ET} + \mathcal{L}_{LR, T}^{CAR} ) \; , \\
S_{RL} & \equiv - \left. \frac{\Delta \mu_{RL}}{e \Delta T_{RL}} \right|_0 = R_{RL, RL} ( \mathcal{L}_{LR, T}^{ET} + \mathcal{L}_{LR, T}^{CAR} ) \nonumber \\ \label{eq:L-float-RL}
& = S_{RS} - S_{LS} \; .
\end{align}
Symbol $(\ldots)|_0$ indicates that we treat also the superconducting reservoir as the floating electrode which implies $J_{RS} = 0 = J_{LS}$~\cite{mazza2014} (Appendix~\ref{app:currhyb}). To simplify the notation we also use the abbreviation $S_{ij} \equiv S_{ij, RL}$.

The local Seebeck coefficient $S_{RL}$ is a linear combination of the non-local thermopowers $S_{LS}$ and $S_{RS}$ which obey the relation $S_{LS} / S_{RS} = R_{LS, RL} / R_{RS, RL}$. It means that in our system only one Seebeck coefficient is independent. In the wide band limit, \textit{i.e.} assuming energy independent couplings $\Gamma_i$, the CAR processes do not enter the thermopower except \textit{via} resistances. This manifests the electron-hole symmetry in the system and formally causes the symmetry of the integrand \eqref{LTcal} with respect to $E$, leading to $\mathcal{L}_{LR, T}^{CAR} = 0$. For this reason we can focus on the thermopower $S_{RL}$ as the other (non-local) thermopowers $S_{RS}$ and $S_{LS}$ can be obtained from $S_{RL}$ [see Eqs.~(\ref{eq:L-float-LS})-(\ref{eq:L-float-RL})] using relations
\begin{equation} \label{eq:srl}
S_{RL} = -\frac{\Gamma_N}{\Gamma_R} S_{LS} = \frac{\Gamma_N}{\Gamma_L} S_{RS} \; ,
\end{equation}
where $\Gamma_N = \Gamma_L + \Gamma_R$. After some algebra we get
\begin{align}
S_{RL} & = \frac{1}{e} \frac{\mathcal{L}_{LR, T}^{ET}}{\mathcal{L}_{LR, \mu}^{ET} + \mathcal{L}_{LR, \mu}^{CAR}} \label{eq:SRLeq} \\
& = \frac{1}{eT} \frac{\int dE \, E \, | G_{11}^r (E)|^2 ( -\frac{\partial f}{\partial E} )}{\int dE \, [ |G_{11}^r (E)|^2 + |G_{12}^r (E)|^2 ](-\frac{\partial f}{\partial E})} \; . \nonumber
\end{align}
In comparison to the two-electrode case the result (\ref{eq:SRLeq}) differs by the additional Andreev reflection term $\mathcal{L}_{LR, \mu}^{CAR}$ appearing in the denominator.

\textit{Measurements} -- We can obtain information about $\mathcal{L}_{LR, T}^{ET} + \mathcal{L}_{LR, T}^{CAR}$ only from the two-step measurements. In the first step one should measure (under isothermal conditions) the resistance $R_{LS, RL}$, $R_{RS, RL}$ or $R_{RL, RL}$ and in the second step (assuming open circuit conditions) the non-local thermopowers $S_{LS}$, $S_{RS}$ or the local thermopower $S_{RL}$ [see Eqs.~(\ref{eq:L-float-LS})-(\ref{eq:L-float-RL})] has to be measured, respectively. The linear coefficient $\mathcal{L}_{LR, T}^{ET} + \mathcal{L}_{LR, T}^{CAR}$ reads
\begin{equation}
\mathcal{L}_{LR, T}^{ET} + \mathcal{L}_{LR, T}^{CAR} = \frac{S_{LS}}{R_{LS, RL}} = \frac{S_{RS}}{R_{RS, RL}} = \frac{S_{RL}}{R_{RL, RL}} \; .
\end{equation}
In the next sections we analyze dependence of $S_{RL}$ on the gate voltage $\varepsilon_0$, considering the high and low temperature regions at various couplings $\Gamma_S$. We study separately the non-interacting ($U = 0$) and the interacting ($U \neq 0$) cases.

\subsection{Non-interacting quantum dot}

\begin{figure} 
\includegraphics[width=0.95\linewidth,clip]{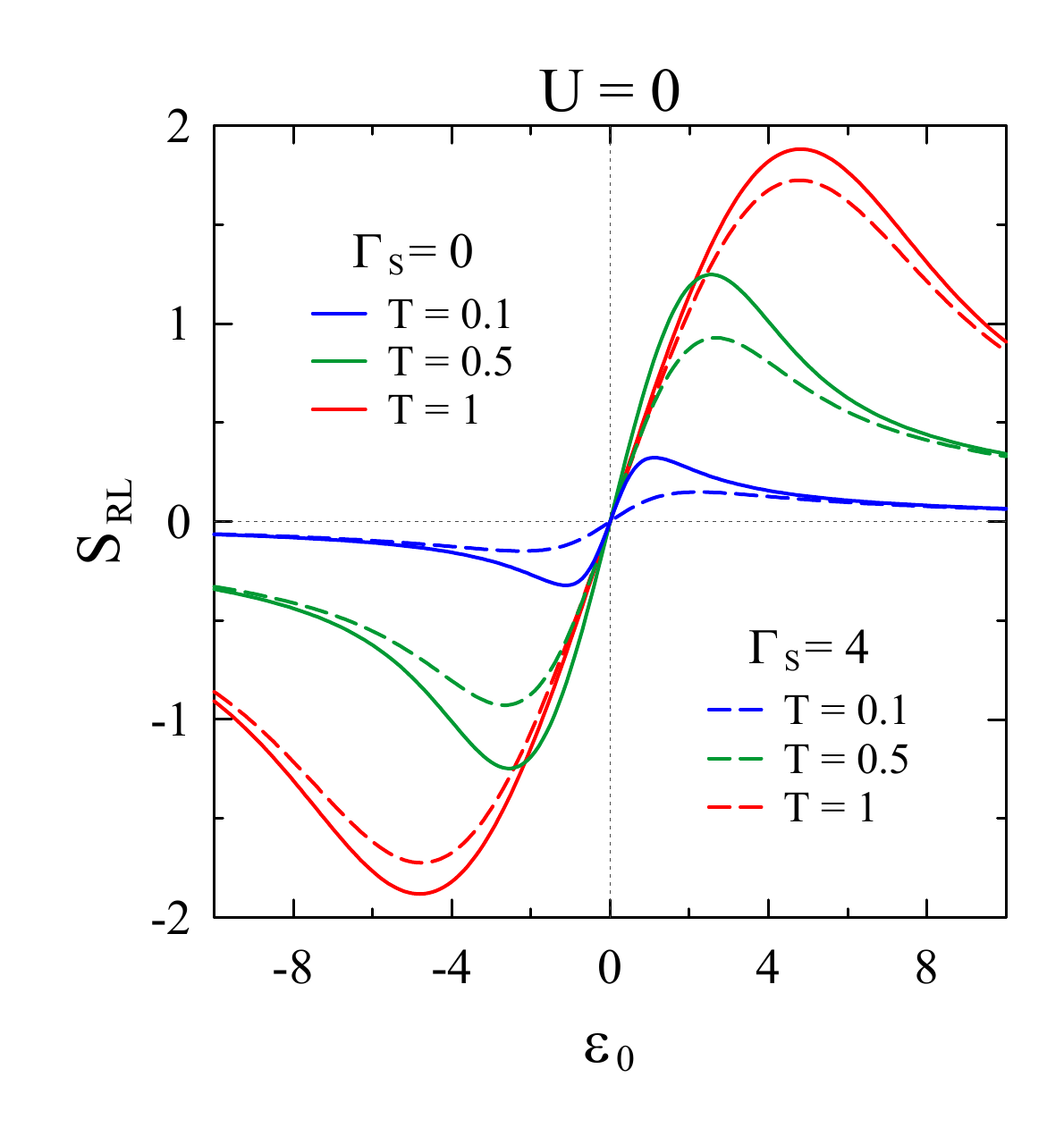}
\caption{\label{fig:Suu0} (Color online) The Seebeck coefficient $S_{RL}$ in units $k_B / e$ as a function of the QD level position $\varepsilon_0$ for: $k_B T / \Gamma_L = 0.1$ (blue lines), $k_B T / \Gamma_L = 0.5$ (green lines), and $k_B T / \Gamma_L = 1$ (red lines). Results are obtained for the uncorrelated QD ($U = 0$) symmetrically coupled to \textit{L} and \textit{R} electrodes ($\Gamma_R / \Gamma_L = 1$) assuming $\Gamma_S = 0$ (solid lines) and $\Gamma_S / \Gamma_L = 4$ (dashed lines). $\Gamma_L$ is taken as unity in calculations.}
\end{figure}

We first study the non-interacting quantum dot. In this limit it is even possible to obtain analytic results, but we prefer to concentrate on the physics of the problem~\cite{mani2009,turek2002} in the three-terminal configuration. Fig.~\ref{fig:Suu0} shows the thermopower $S_{RL}$ as a function of the gate voltage (\textit{via} the QD level $\varepsilon_0$) at various temperatures for symmetric coupling $\Gamma_R / \Gamma_L = 1$ and $\Gamma_S = 0$ (solid lines) or $\Gamma_S / \Gamma_L = 4$ (dashed curves). We have assumed the large superconducting energy gap limit ($\Delta \rightarrow \infty$), when the proximity effect yields the Andreev bound states~\cite{michalek2013} formed at $E = \pm \sqrt{\varepsilon_0^2 + (\Gamma_S / 2)^2}$. Their line-broadening $\Gamma_N = \Gamma_L + \Gamma_R$ depends on the couplings to \textit{L} and \textit{R} electrodes. For small coupling $\Gamma_S \ll \Gamma_N$, these states practically merge into a single broad feature centered at zero energy, while for larger $\Gamma_S \gg \Gamma_N$ the separate peaks are seen~\cite{michalek2013}. Such an electronic spectrum indirectly affects the thermopower.

\textit{High temperature limit} -- For temperatures $k_B T \gtrsim \Gamma_L $ and the QD level $\varepsilon_0$ close to the Fermi energy the system is in the sequential tunneling regime. For weak coupling $\Gamma_S$ the Andreev reflection is suppressed and electrons can flow between normal electrodes through the QD level. Since large temperatures imply $f_L \simeq f_R$ we obtain $S_{RL} \propto \varepsilon_0 / T$. For $| \varepsilon_0 | \gg ( \Gamma_i, k_B T )$ the sequential tunneling processes are exponentially suppressed and the cotunneling becomes the main contribution channel. In this regime the thermopower shows the characteristic metallic dependence $S_{RL} \propto T / \varepsilon_0$~[\onlinecite{turek2002}]. With increasing coupling $\Gamma_S$ the anomalous Andreev reflection processes affect the denominator of (\ref{eq:SRLeq}), slightly suppressing $S_{RL}$.

\textit{Low temperature limit} -- At low temperatures $k_B T \ll \Gamma_i$ and assuming weak energy dependence of the transmittance $T^{\kappa} (E)$ one can evaluate the coefficients $\mathcal{L}_{\alpha \beta, \gamma}^\kappa$ by means of the Sommerfeld expansion. In the case of uncorrelated QD the thermopower $S_{RL}$ simplifies to
\begin{equation} \label{eq:SRLSomm}
S_{RL} = \frac{2}{e} \frac{\pi^2}{3} k_B^2 T \frac{\varepsilon_0}{\varepsilon_0^2 + \Gamma_N^2/4 + \Gamma_S^2/4} \; .
\end{equation}
The thermopower (\ref{eq:SRLSomm}) has a metallic-like character and its magnitude decreases with increasing the coupling $\Gamma_S$, as could have been inferred from \eqref{eq:SRLeq}.

%
%

\subsection{Correlation effects}

%
\begin{figure*} 
\includegraphics[width=0.82\linewidth,clip]{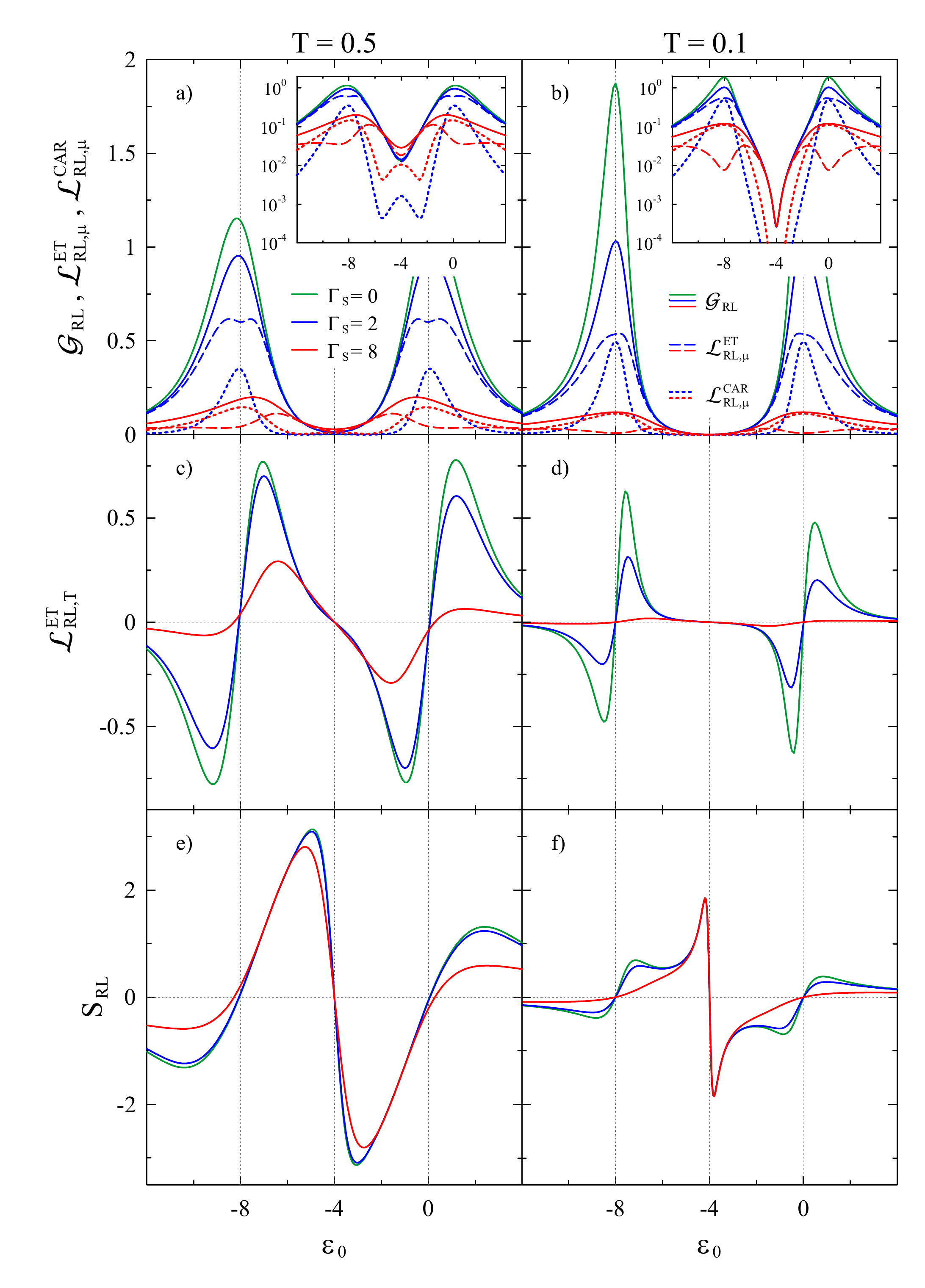}
\caption{\label{fig:Suutt}(Color online) The gate voltage dependence of: a) and b) the total conductance $\mathcal{G}_{RL} = e ( \mathcal{L}_{RL, \mu}^{ET} + \mathcal{L}_{RL, \mu}^{CAR} )$ (solid lines), $\mathcal{L}_{RL, \mu}^{ET}$ (dashed lines) and $\mathcal{L}_{RL, \mu}^{CAR}$ (dotted lines) (insets show the results in a logarithmic scale); c) and d) the linear coefficient $\mathcal{L}_{RL, T}^{ET}$; e) and f) the Seebeck coefficient $S_{RL}$ (in units $k_B / e$) of the correlated QD with Coulomb potential $U / \Gamma_L = 8$ for the symmetric coupling $\Gamma_R / \Gamma_L = 1$ and several couplings to the superconducting electrode $\Gamma_S / \Gamma_L = \{ 0, 2 ,8 \}$ (green, blue and red lines) in the large $k_B T / \Gamma_L = 0.5$ (left column) or low $k_B T / \Gamma_L = 0.1$ (right column) temperature.}
\end{figure*}
%

Strong repulsion between the opposite-spin electrons can induce the Coulomb blockade (CB) and, at sufficiently low temperatures, may produce the Kondo effect~\cite{hewson2007}. The correlations can thus indirectly affect the thermoelectric properties, as has been already shown in the two-terminal~\cite{wierzbicki2011,trocha2012} and the three-terminal~\cite{wysokinski2012} systems. Furthermore, in hybrid structures with the superconducting leads, the Coulomb potential $U$ has qualitative effect on the Andreev bound states~\cite{wysokinski2012}. In consequence the interactions may substantially suppress the conductances of the three-terminal system (Fig.~\ref{fig1}), except only in the vicinity of the Andreev bound states~\cite{michalek2013}.

Here we study influence of the Coulomb interactions on the thermopower $S_{RL}$. For this purpose we apply the Hubbard I approximation, which qualitatively captures the Coulomb blockade effect (see the Appendix~\ref{app:model} for technical details). In the large $\Delta$ limit, the subgap spectrum consists of four Andreev bound states, determined by poles of the Green function $G_{11}^r (E)$. For $\Gamma_L, \Gamma_R \rightarrow 0$ they are located at energies~\cite{michalek2013}
\begin{equation} \label{eq:AR}
E_{\lambda, \lambda'} = \frac{\lambda}{\sqrt{2}} \sqrt{\varepsilon_0^2 + \varepsilon_U^2 + \Gamma_S^2 / 4 + \lambda' \delta} \; ,
\end{equation}
where $\delta = \sqrt{( \varepsilon_0^2 + \varepsilon_U^2 + \Gamma_S^2 / 4 )^2 - ( \Gamma_S^2 \varepsilon_n^2 + 4 \varepsilon_0^2 \varepsilon_U^2 )^2}$, $\varepsilon_U = \varepsilon_0 + U$, $\varepsilon_n = \varepsilon_0 + ( 1 - n / 2 ) U$, $\lambda, \lambda' = \pm 1$ and $n$ is an average electron number of the QD.

The in-gap features (\ref{eq:AR}) represent the quasiparticle excitations between the singly occupied states $\left| \sigma \right>$ and two coherent superpositions of the empty $\left| 0 \right>$ and doubly occupied configurations $\left| \uparrow \downarrow \right>$. Their position, line-broadening and spectral weights can be tuned by the gate voltage (\textit{via} $\varepsilon_0$) and by the coupling $\Gamma_S$~\cite{michalek2013}. The resulting spectrum can be indirectly probed, by measuring the subgap conductance where nontrivial interplay between the normal electron transfer (ET) and anomalous Andreev reflection (DAR and CAR) channels occur~\cite{michalek2015}.
Fig.~\ref{fig:Suutt} shows the thermopower~\eqref{eq:SRLeq} defined as
\begin{equation} \label{tep-cond}
S_{RL} = \frac{\mathcal{L}_{LR, T}^{ET}}{\mathcal{G}_{RL}} \; ,
\end{equation}
as well as its ingredients: total conductance $\mathcal{G}_{RL} = e ( \mathcal{L}_{RL, \mu}^{ET} + \mathcal{L}_{RL, \mu}^{CAR} )$ and thermal coefficient $L_{RL, T}^{ET}$ as a function of the gate voltage $\varepsilon_0$. The results were obtained numerically for the strong Coulomb interaction $U / \Gamma_L = 8$ at both high ($k_B T / \Gamma_L = 0.5$) and low ($k_B T / \Gamma_L = 0.1$) temperatures for symmetric coupling $\Gamma_R / \Gamma_L = 1$ and various coupling to the superconducting electrode $\Gamma_S / \Gamma_L = \{ 0, 2, 8 \}$.

\textit{Conductance} -- Both components of $\mathcal{G}_{RL}$ are very sensitive to the pairing correlations induced by the proximity effect. Since~\eqref{Lmucal} is a quantitative measure of the subgap spectrum, the Andreev bound states~\eqref{eq:AR} show up in the ET component $\mathcal{L}_{RL, \mu}^{ET}$. The component $\mathcal{L}_{RL, \mu}^{CAR}$ also reveals maxima around the same Andreev bound states~\eqref{eq:AR} but with different amplitude encoded in the off-diagonal terms of the Nambu Green's function~\cite{michalek2015}.

When $\Gamma_S < \Gamma_N$ the ET processes dominate over CAR scattering in entire gate voltage regime~\cite{michalek2015a}. However, for $\Gamma_S > \Gamma_N$ the situation is different. Inside the Coulomb blockade (CB) region, \textit{i.e.} between the inner Andreev bound states $E_{-, - }$ and $E_{+, -}$, the ET domination is still visible because the Coulomb repulsion suppresses the CAR more efficiently than the electron transfer~\cite{michalek2015}. The extent of the CB region is governed by the coupling $\Gamma_S$ and it shrinks upon increasing the on-dot pairing~\cite{michalek2015a}. On the other hand, outside the CB region the CAR processes dominate over the ET processes in a vicinity of the inner Andreev bound states. For larger $|\varepsilon_{0}|$ transport is again dominated by the ET processes (because the proximity induced pairing is very weak). Change of the dominant transport channel causes the negative non-local conductance already observed in planar systems~\cite{russo2005,cadden2006,brauer2010,webb2012} but not in tunnel structures as considered here. This effect, resulting from a competition between the crossed Andreev reflections and the direct electron tunneling involving normal electrodes, can be detected by measuring the non-local resistance $R_{RS, LS} \propto \mathcal{L}_{LR}^{ET} - \mathcal{L}_{LR}^{CAR}$ as a function of the gate potential in the three-terminal hybrid device with QD~\cite{michalek2015}.

Inset of Fig.~\ref{fig:Suutt}~b) shows that, at low temperature $k_B T \ll \Gamma_L$, the total conductance $\mathcal{G}_{RL}$ exponentially diminishes near $\varepsilon_0 = - U / 2$. This effect is caused by destructive interference between the electrons tunnelled through the inner Andreev bound states in some analogy to the normal two-level system~\cite{nakanishi2007}.

\textit{Thermal coefficient} -- Coefficient $\mathcal{L}_{RL, T}^{ET}$ in general behaves similar to $\mathcal{G}_{RL}$ \textit{i.e.} it diminishes with an increase of coupling $\Gamma_S$ everywhere except in the vicinity of the electron-hole symmetry point $\varepsilon_0 = - U / 2$. In this region $\mathcal{L}_{RL, T}^{ET}$ rather weakly depends on $\Gamma_S$. At low temperatures, when the Sommerfeld approximation can be applied, $\mathcal{L}_{RL, T}^{ET} \propto \partial\mathcal{L}_{RL, \mu}^{ET} / \partial E$ and (for $\Gamma_S = 0$) one recovers the popular Mott formula for thermopower~\cite{nakanishi2007} \textit{i.e.} $S \propto \partial \ln \mathcal{L}_\mu / \partial E$. Additionally, the inner peaks are more transparent then the outer ones due to the asymmetry caused by the Coulomb blockade. This effect is gradually suppressed at larger temperatures.

\textit{Thermpower} -- The Seebeck coefficient $S_{RL}$ in a low temperature $k_B T / \Gamma_L = 0.1$ regime is shown in Fig.~\ref{fig:Suutt}~f). When $\varepsilon_0 \approx - U$ and $\varepsilon_0 \approx 0$ the shape of the $S_{RL}$ resembles behavior obtained in the non-interacting case (Fig.~\ref{fig:Suu0}). As before these $S_{RL}$ peaks (we call them ``normal'' peaks) are a result of a competition between sequential tunnelling (when $S_{RL} \sim \varepsilon_0 / T$) that dominates for $\varepsilon_0$ which are close to conductance peaks and cotunnelling ($S_{RL} \sim T / \varepsilon_0$) that dominates in the remaining region. A corresponding sign change of the $S_{RL}$ coincide with the changeover between dominant transport carriers from the electrons to holes.

The Coulomb interaction strongly affects the thermopower in a valley between the conductance peaks. In particular an additional sharp structure appears close to the electron/hole symmetry point $\varepsilon_0 = - U / 2$. This effect is related to the Fano resonance~\cite{bulka2001,baranski2001} and is a signature of the destructive interference between the activated QD levels. It leads to reduction of the total transmission to zero and, as a consequence, to a reduction of $\mathcal{G}_{RL}$ and $\mathcal{L}_{RL, T}^{ET}$~\cite{nakanishi2007}. Since the reduction of $\mathcal{G}_{RL} \propto (\varepsilon_0 - U / 2)^2$ is stronger than $\mathcal{L}_{RL, T} \propto ( \varepsilon_0 - U / 2)$ the $S_{RL}$ is enhanced.

For $\Gamma_S > \Gamma_N$ the CAR and the ET processes compete~\cite{michalek2013}, suppressing $\mathcal{L}_{RL, \mu}^{ET}$ and $\mathcal{L}_{RL, T}^{ET}$ in the vicinity of conductance peaks. This leads to a suppression of ``normal'' $S_{RL}$ peaks. Since $\Gamma_S$ has week influence on transport coefficients in the vicinity of the electron-hole symmetry point no change in the sharp structure can be observed. This behaviour should be contrasted with the one observed for the three-terminal system with the normal electrodes laterally connected through QD, where the thermopower in CB region is strongly suppressed due to phase randomization induced by the third electrode~[\onlinecite{nakanishi2007}].

\begin{figure} 
\includegraphics[width=\linewidth,clip]{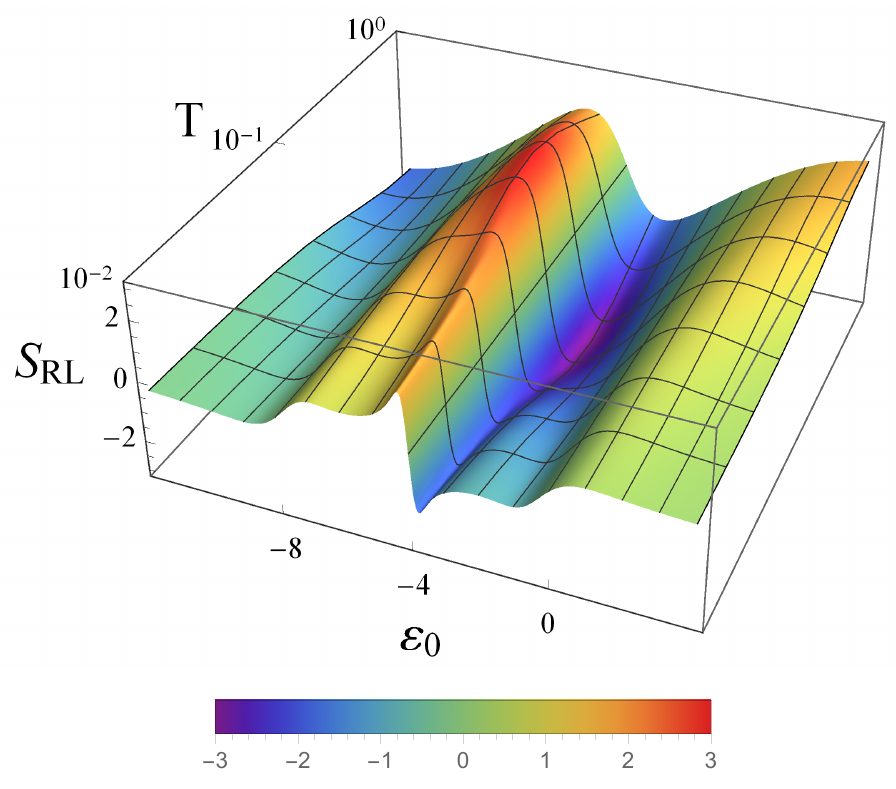}
\caption{\label{fig:Stemp}(Color online) The Seebeck coefficient $S_{RL}$ (in units of $k_B / e$) map in the ($\varepsilon_0, T$) space for $U / \Gamma_L = 8$, $\Gamma_R / \Gamma_L = 1$ and $\Gamma_S / \Gamma_L = 2$.}
\end{figure}

The Seebeck coefficient $S_{RL}$ in a high temperature regime and its temperature evolution as a function of $\varepsilon_0$ is shown in Figs.~\ref{fig:Suutt}~e) and~\ref{fig:Stemp}, respectively. The non-monotonic temperature dependence of $S_{RL}$ in the Coulomb blockade region is a result of an interplay between the cotunnelling (which dominates at low temperatures) and the sequential tunnelling transport (which dominates at high temperatures). With an increasing temperature $T$, the shape of the $S_{RL}$ evolves. The peaks observed at low temperature merge together and they become hardly distinguishable at larger temperatures. Their combined amplitude achieves maximum at $k_B T / \Gamma_L \approx 0.5$ which is almost twice as high as in the corresponding non-interaction case ($U = 0$). The amplitude achieves large value up to $\sim 270 \mu V / K$, comparable with thermopower in a transistor-like structure with strong interface spin polarizations~\cite{machon2014}. For very large temperature one observes a well known (and experimentally verified) saw-tooth shape of the thermopower~\cite{costi2010,beenakker1992,dzurak1997} predicted by a sequential tunneling, where $S_{RL} \sim \varepsilon_0 / T$. Similarly to the low temperature case the magnitude of thermopower only slightly diminishes with an increasing $\Gamma_S$ in the CB region.

\textit{Asymmetric coupling} -- So far we have presented results for the symmetric couplings $\Gamma_R / \Gamma_L = 1$. Since any left-right asymmetry $\Gamma_R \neq \Gamma_L$ does not break the electron-hole symmetry one would observe qualitatively the same characteristics, with some quantitative changes only in the magnitude of $S_{RL}$. Asymmetry of the couplings has no effect on the position of the sharp structure related to the Fano resonance either. This should be contrasted with the properties reported for the systems with the assisted hopping processes, where the average charge and the phase shift are linked through the Friedel sum-rule~\cite{tooski2014}.

Finally let us remark, that the similar sharp structure does also appear in the system with two-level QD attached to two normal electrodes and one s-wave superconductor~\cite{valentini2015}. The situation described in Ref.~[\onlinecite{valentini2015}] (where one level is pinned at the Fermi energy and only the other one varies with the gate voltage) seems to be hardly realistic.

%
%

\section{\label{sec:other_definitions}Other definitions of thermopower}

There have been discussed various proposals~\cite{sanchez2011,wysokinski2012,mazza2014,machon2013}, generalizing the two-terminal thermopower (\ref{def-tep}) to multi-terminal structures. The temperature difference $\Delta T_j$ and the bias $\Delta V_i$ may be measured (with respect to any reference values) either at the same (or different) electrode(s), therefore the local (non-local) Seebeck coefficient $S_{ij}$ can be generally defined as
\begin{equation} \label{S-def-3t}
S_{ij} = - \left. \frac{\Delta V_i}{\Delta T_j} \right|_{\{ cond \}} \; ,
\end{equation}
where $\{ cond \}$ refers to the experimental constraints and is another subject of arbitrariness.

In this section we would like to compare various definitions existing in the literature with our experimentally feasible proposal.

\textit{``Mazza's'' definition} -- Mazza \textit{et al.}~\cite{mazza2014} have recently discussed question how to define the non-local thermopower and pointed out conditions under which the non-local effects can be observed in the linear response transport through a three-terminal system. Assuming a reservoir 3 as a reference with $\mu_3 = \mu$ and $T_3 = T$, taking into account the charge and the energy conservation, the authors have expressed a particle $J_i^N$ and a heat $J_i^Q$ currents flowing from reservoir $i = \{ 1, 2 \}$ through the Onsager matrix of elements $L_{ij}$
\begin{equation}\label{mazza-curr}
\def\arraystretch{1.5}
\left(
 \begin{array}{c}
 \displaystyle{J_1^N} \\
 \displaystyle{J_1^Q} \\
 \displaystyle{J_2^N} \\
 \displaystyle{J_2^Q} \\
 \end{array}
\right)
=
\left(
 \begin{array}{cccc}
 L_{11} & L_{12} & L_{13} & L_{14} \\
 L_{21} & L_{22} & L_{23} & L_{24} \\
 L_{31} & L_{32} & L_{33} & L_{34} \\
 L_{41} & L_{42} & L_{43} & L_{44} \\
 \end{array}
\right)
\left(
 \begin{array}{c}
 X_1^\mu \\
 X_1^T \\
 X_2^\mu \\
 X_2^T \\
 \end{array}
\right) \; ,
\end{equation}
where $X_i^\mu = \Delta \mu_i / T$, $X_i^T = \Delta T_i / T^2$ and $\Delta \mu_i = \mu_i - \mu$, $\Delta T_i = T_i - T$.

\begin{figure}
\includegraphics[width=0.98\linewidth,clip]{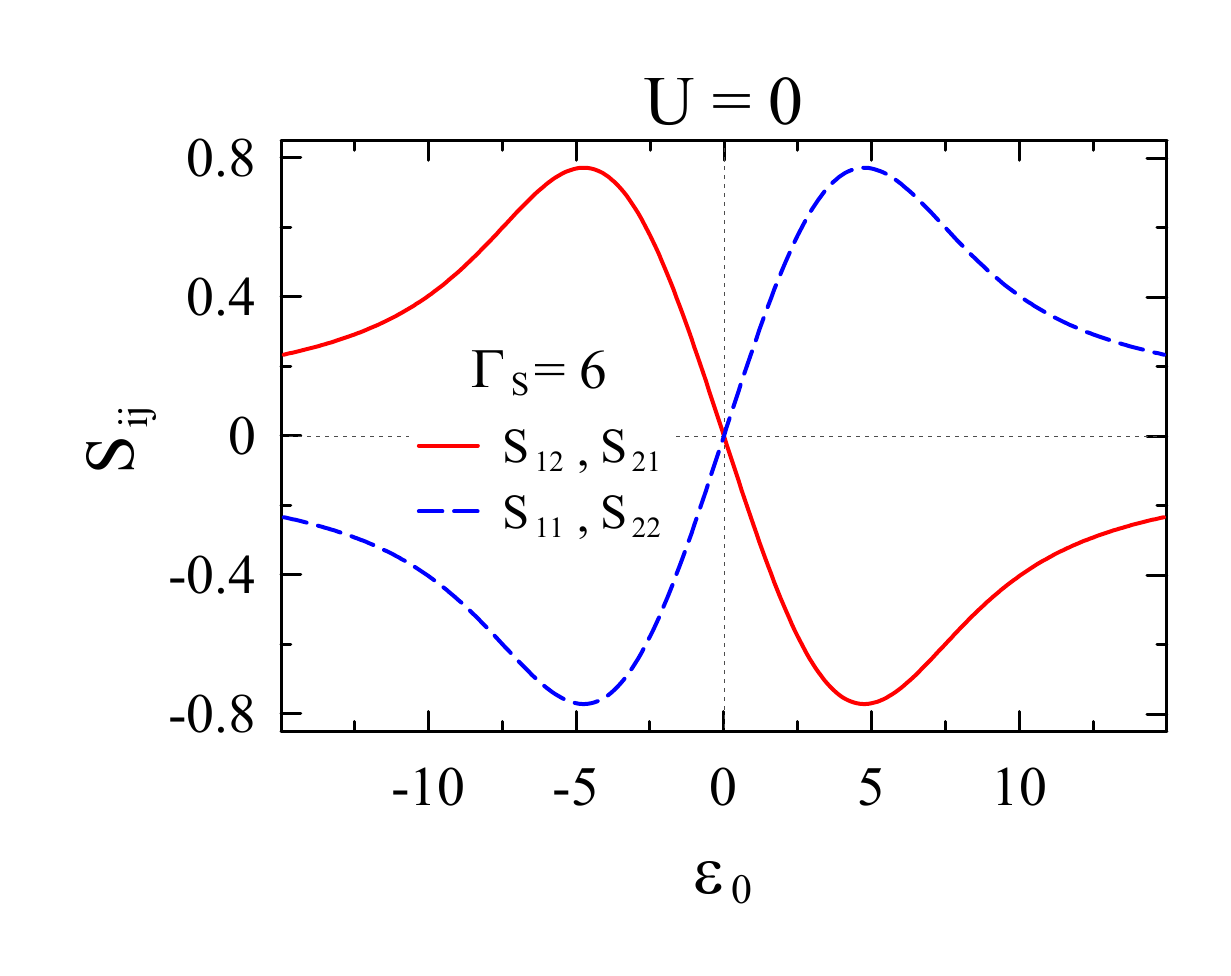}
\caption{\label{fig:teps-mazza}(Color online) Gate voltage dependence of the Seebeck coefficients (in units $k_B / e$) as defined in equations (\ref{teps-1})-(\ref{teps-4}). Notice the difference between the local $S_{11} = S_{22}$ (blue dashed line) and non-local $S_{12} = S_{21}$ (red solid line) Seebeck coefficients. The results are obtained for $U = 0$, $\Gamma_S / \Gamma_L = 6$, $\Gamma_R / \Gamma_L = 1$ and $k_B T / \Gamma_L = 1$.}
\end{figure}

Then the Seebeck coefficients have been introduced by imposing vanishing particle currents in all leads, $J_1^N = J_2^N = J_3^N = 0$, together with $\Delta T_2 = 0$ and $\Delta T_1 \neq 0$ or $\Delta T_1 = 0$ and $\Delta T_2 \neq 0$. In the linear response regime the authors~\cite{mazza2014} have found the following thermopowers
\begin{align} \label{teps-1}
& S_{11} = - \left. \frac{\Delta \mu_1}{e \Delta T_1} \right|_{\Delta T_2 = 0} = \frac{1}{eT} \frac{L_{13} L_{23} - L_{12} L_{33}}{L_{13} L_{13} - L_{11} L_{33}} \; , \\ \label{teps-2}
& S_{12} = - \left. \frac{\Delta \mu_1}{e \Delta T_2} \right|_{\Delta T_1 = 0} = \frac{1}{eT} \frac{L_{13} L_{43} - L_{14} L_{33}}{L_{13} L_{13} - L_{11} L_{33}} \; , \\ \label{teps-3}
& S_{21} = - \left. \frac{\Delta \mu_2}{e \Delta T_1} \right|_{\Delta T_2 = 0} = \frac{1}{eT} \frac{L_{12} L_{13} - L_{11} L_{23}}{L_{13} L_{13} - L_{11} L_{33}} \; , \\ \label{teps-4}
& S_{22} = - \left. \frac{\Delta \mu_2}{e \Delta T_2} \right|_{\Delta T_1 = 0} = \frac{1}{eT} \frac{L_{13} L_{14} - L_{11} L_{34}}{L_{13} L_{13} - L_{11} L_{33}} \; .
\end{align}
Fig.~\ref{fig:teps-mazza} shows the gate voltage dependence of the Seebeck coefficients (\ref{teps-1})-(\ref{teps-4}) obtained within the scheme described in~[\onlinecite{mazza2014}]. The local and non-local coefficients obey the symmetry relation $S_{11 (22)} (\varepsilon_0) = - S_{12 (21)} (\varepsilon_0)$ and under optimal conditions they approach the large value of the order of $k_B / e \approx 86.17 \mu V/K$.

Using definition (\ref{def:theropower}) and assuming that $\Delta T_{RS} = 0$ or $\Delta T_{LS} = 0$ one obtains the Seebeck coefficients corresponding to (\ref{teps-1})-(\ref{teps-4})
\begin{align}
S_{LS, LS} & = - \left. \frac{\Delta \mu_{LS}}{e \Delta T_{LS}} \right|_{\Delta T_{RS} = 0} = - \frac{s_{LS, LR}}{e} - \frac{s_{LS, LS}}{e} \; , \\
S_{LS, RS} & = - \left. \frac{\Delta \mu_{LS}}{e \Delta T_{RS}} \right|_{\Delta T_{LS} = 0} = \frac{s_{LS, LR}}{e} - \frac{s_{LS, RS}}{e} \; , \\
S_{RS, LS} & = - \left. \frac{\Delta \mu_{RS}}{e \Delta T_{LS}} \right|_{\Delta T_{RS} = 0} = - \frac{s_{RS, LR}}{e} - \frac{s_{RS, LS}}{e} \; , \\
S_{RS, RS} & = - \left. \frac{\Delta \mu_{RS}}{e \Delta T_{RS}} \right|_{\Delta T_{LS} = 0} = \frac{s_{RS, LR}}{e} - \frac{s_{RS, RS}}{e} \; .
\end{align}
Here we recall that $s_{mn, ij} = - e R_{mn, ij} \mathcal{L}_{ij, T}$.

\textit{``Machon's'' definition} -- Machon \textit{et al.}~\cite{machon2013} have studied charge transport in a system which consists of one superconducting (S) and two ferromagnetic contacts. The authors have proposed (in addition to the aforementioned Mazza definition) two additional constraints to define and analyze non-local Seebeck coefficients.

In the first case they assumed that the charge current through contact 1 into superconductor vanishes ($I_1^q = 0$), $\Delta T_1 = T_1 - T_S = 0$ and $\Delta V_2 = V_2 - V_S = 0$. In our notation these constraints read $J_L = 0$, $\Delta T_{LS} = 0$ and $\Delta \mu_{RS} = 0$ so the non-local Seebeck coefficient defined by (\ref{def:theropower}) takes the form
\begin{equation}
S_{LS, LR} = - \frac{\Delta \mu _{LS}}{e \Delta T_{LR}} = \frac{1}{e} \frac{\mathcal{L}_{LR, T}}{\mathcal{L}_{LR, \mu} + \mathcal{L}_{LS, \mu}} \; .
\end{equation}

In the second case the authors required the vanishing sum of charge currents $I_1^q + I_2^q = 0$ with $\Delta T_1 = 0$ and $\Delta V_2 = 0$. The non-local thermopower calculated from (\ref{def:theropower}) and equivalent to the second proposition by Machon \textit{et al.}~\cite{machon2013} reads
\begin{equation}
S_{LS, LR} = - \frac{\Delta \mu_{LS}}{e \Delta T_{LR}} = - \frac{1}{e} \frac{\mathcal{L}_{RS, T}}{\mathcal{L}_{LS, \mu}} \; .
\end{equation}
It is easy to check that in both cases $S_{LS, LR} = S_{LR, LR} = - S_{LR, RS} = - S_{LS, RS}$.

\textit{Sanchez's and Serra's definition} -- Considering the open circuit conditions, treating $T_S$ as a reference temperature and describing $T_L$ and $T_R$ in terms of (commonly employed in measurements) symmetric deviation from this temperature: $T_L = T_S + \Delta T_{LR} / 2$, $T_R = T_S - \Delta T_{LR} / 2$ one recovers from (\ref{def:theropower}) the thermopower proposed by Sanchez and Serra~\cite{sanchez2011}
\begin{align}
& S_{LR, LR} = - \frac{1}{\mathcal{G}_{LR, LR}} \nonumber \\
& \times \left[ \mathcal{L}_{LR, T} + \frac{\mathcal{L}_{LS, \mu} \mathcal{L}_{RS, T} + \mathcal{L}_{RS, \mu} \mathcal{L}_{LS, T}}{2 ( \mathcal{L}_{LS, \mu} + \mathcal{L}_{RS, \mu} ) } \right] \; .
\end{align}

In the case $\mathcal{L}_{SL, \mu} \rightarrow 0$ and $\mathcal{L}_{RS, \mu} \rightarrow 0$ (when the superconducting probe is detached) the usual two-terminal thermopower $S = - \mathcal{L}_{LR, T} / e \mathcal{L}_{LR, \mu}$ is reproduced. The adiabatic situation, when temperature of the third electrode is determined self-consistently by requiring vanishing of the heat current has been discussed in Refs [\onlinecite{sanchez2011,wysokinski2016}].

Still another constraint has been adopted in the paper [\onlinecite{wysokinski2012}]. The author has studied the three-terminal system with quantum dot coupled to the normal (N) metal, the superconductor (S) and the ferromagnet (F). To calculate the (charge and spin) thermopowers it has been assumed that the temperature bias was applied to the normal electrode only, $\Delta T_N \ne 0$ and the voltage $V_N$ was required to compensate the current in that electrode $J_N=0$. The charge Seebeck coefficient has been defined in analogy to the two-terminal case as $S=-(V_N/\Delta T_N)_{J_N=0}$.

Such plethora of the possible experimental constraints shows that the thermopower might completely differ from case to case. Therefore we have proposed the two-step protocol for the multi-terminal systems by generalizing the B\"{u}ttiker formalism of the local and non-local resistances~\cite{buttiker1986} on the thermopowers.

%
%

\section{\label{sec:summary}Summary and conclusions}

We have investigated the thermoelectric effects in the three-terminal devices and proposed the original two-step protocol for measuring the transport coefficients. We believe that this algorithm (explained in Sec.~\ref{sec:normal}) and its application to three-terminal device with the superconducting electrode (Sec.~\ref{sec:hybrid}) would be stimulating for the experimentalists and theoreticians as well.

This method extends the approach~\cite{buttiker1986} of the local/non-local resistances $R_{kl, mn}$ to the thermopowers $S_{mn, ij}$. Contrary to the two-terminal situations, no unique definition of the Seebeck coefficient can be formulated for multi-terminal junctions. We have identified a set of experimentally consistent constraints and shown how the measured local and non-local resistances and Seebeck coefficients are related to the linear kinetic coefficients. Similarly to the two-terminal case, the thermopower probes the energy dependence of the transmission function and it vanishes in the particle-hole symmetric case.

We have also studied in detail the thermopower of three-terminal hybrid structure, comprising the quantum dot (QD) coupled to one superconducting and two metallic electrodes. Usually the superconducting electrodes have rather negligible influence on the thermopower, due to the particle-hole symmetry present in the subgap regime. Our results show that indeed there is no direct contribution to the thermopower originating from the Andreev reflections. Formally this can be seen from Eq.~\eqref{eq:SRLeq}, where the crossed Andreev processes appear only in the denominator. This means that the superconducting proximity effect has a direct influence only on the resistances (or conductances), whereas the thermopowers are affected indirectly.

The pronounced thermoelectric effects have been previously reported for the ferromagnet-superconductor junctions due to the spin-splitting (Zeeman) field~\cite{oazeta2014}, where the particle-hole symmetry was broken. In our setup (Fig.~\ref{fig1}) the non-local thermopower is comparable to the local one, due to the Andreev bound-states that substantially affect the subgap differential conductance \cite{michalek2015}. Their indirect influence on the thermopower might be promising for innovative applications, \textit{e.g.} in the energy harvesting, nanothermometers, cooling nanodevices, \textit{etc.}

\begin{acknowledgments}
This work is supported by the National Science Centre in Poland by the projects DEC-2012/05/B/ST3/03208 (GM, MU, BRB) and DEC-2014/13/B/ST3/04451 (TD, KIW).
\end{acknowledgments}

\appendix

\section{\label{app:model}Microscopic model for the three- terminal structure with QD}

The system displayed in Fig.\ \ref{fig1} can be described by the Anderson-impurity Hamiltonian~\cite{michalek2013}
\begin{equation} \label{eq-Ham}
H = H_{QD} + \sum_{\alpha} H_\alpha + H_T \; ,
\end{equation}
where $\alpha = L, R, S$. $H_{QD}$ describes the quantum dot
\begin{equation} \label{eq-Ham-QD}
H_{QD} = \varepsilon_0 \sum_\sigma d_\sigma^\dag d_\sigma + U n_\uparrow n_\downarrow \; ,
\end{equation}
where $\varepsilon_0$ is the single-particle energy level, $d_\sigma^\dag$ ($d_\sigma$) denotes creation (annihilation) operator of the dot electron with spin $\sigma$, $n_\sigma \equiv d_\sigma^\dag d_\sigma$, and $U$ is the Coulomb interaction between two opposite spin electrons. The second term in (\ref{eq-Ham}) refers to electrons in the leads
\begin{equation}
H_\alpha = \sum_{k, \sigma} \varepsilon_{\alpha k} c_{\alpha k \sigma}^\dag c_{\alpha k \sigma} \; ,
\end{equation}
where $c_{\alpha k \sigma}^\dag$ ($c_{\alpha k \sigma}$) is the creation (annihilation) operator of electron with spin $\sigma$ and momentum $k$ in the electrode $\alpha = \{ L, R, S \}$.

The above expression refers to all leads being normal metallic. If one of the electrodes is superconducting its Hamiltonian has to be supplemented by the part describing the condensate, so the full Hamiltonian of the superconducting electrode is represented by the BCS bi-linear Hamiltonian
\begin{align}
H_S & = \sum_{k, \sigma} \varepsilon_{S k} c_{S k \sigma}^\dag c_{S k \sigma} \nonumber \\
& + \sum_k \left( \Delta c_{S -k \uparrow}^\dag c_{S k \downarrow}^\dag + \Delta^* c_{S k \downarrow} c_{S -k \uparrow} \right) \; ,
\end{align}
assuming the isotropic energy gap $\Delta$. The coupling between QD and external leads (whether superconducting or normal) is given by
\begin{equation}
H_T = \sum_{\alpha, k, \sigma} \left( t_\alpha c_{\alpha k \sigma}^\dag d_\sigma + t_\alpha^* d_\sigma^\dag c_{\alpha k \sigma} \right) \; ,
\end{equation}
where $t_\alpha$ is the hopping integral between QD and the itinerant electrons of $\alpha$ lead. In the wide band limit, the electron and hole transfers between the QD and the leads can be described by the tunneling rate $\Gamma_\alpha = 2 \pi \sum_k |t_\alpha|^2 \delta ( E - \varepsilon_{\alpha k} ) = 2 \pi | t_\alpha |^2 \rho_\alpha$, where $\rho_\alpha$ is the density of states in the $\alpha$ electrode defined for $\varepsilon_{\alpha k}$ spectrum.

The details of the calculations of the currents in the system depend if all electrodes are normal or the superconducting electrode is present. In the former case one uses scalar Green functions within the Keldysh approach to the non-equilibrium transport. On the other hand in hybrid system with one or more superconducting electrodes one needs the retarded Green function $\hat{G}^r (E)$ of the QD in the Nambu spinor representation. In both cases the current can be expressed in terms of retarded and lesser elements of the Keldysh Green function.

In the following we concentrate on the latter case, which is more involved. There are two important interactions which contribute to the self-energy entering the Greens functions of the dot: the Coulomb on-dot interactions and the coupling to the leads. They can be included \textit{via} the Dyson equation
\begin{equation} \label{Dyson}
\hat{G}^r (E) = \hat{g}^r (E) + \hat{g}^r (E) \hat{\Sigma}^r (E) \hat{G}^r (E) \; ,
\end{equation}
where $\hat{g}^r (E)$ corresponds to the isolated and non-interacting QD
\begin{eqnarray} \label{GF_T}
\hat{g}^r (E) =
\def\arraystretch{2}
\left(
 \begin{array}{cc}
    \displaystyle{\frac{1}{E - \varepsilon_0 + i 0^+}} & 0 \\
    0 & \displaystyle{\frac{1}{E + \varepsilon_0 + i 0^+}}
 \end{array}
\right) \; ,
\end{eqnarray}
and $\hat{\Sigma}^r (E)$ is the appropriate self-energy. We can express the matrix Green's function by
\begin{align} \label{G11_U}
G_{11}^r & = \frac{1 / g_{22}^r - \Sigma_{22}^r}{\left( 1 / g_{11}^r - \Sigma_{11}^r \right) \left( 1 / g_{22}^r - \Sigma_{22}^r \right) - \Sigma_{12}^r \Sigma_{21}^r} \; , \\ \label{G12_U}
G_{12}^r & = - \frac{\Sigma_{12}^r}{1 / g_{22}^r - \Sigma_{22}^r} G_{11}^r
\end{align}
The self-energy matrix consists of two contributions
\begin{equation}
\hat{\Sigma}^r = \hat{\Sigma}_T^r + \hat{\Sigma}_U^r \; ,
\end{equation}
where $\hat{\Sigma}_T^r$ accounts for the coupling between QD and the leads and $\hat{\Sigma}_U^r$ stands for the self-energy due to correlations. In the `superconducting atomic limit' (\textit{i.e.} deep inside the superconducting energy gap) the first contribution reads~\cite{tanaka2007}
\begin{equation} \label{sigma0}
\hat{\Sigma}_T^r =
\def\arraystretch{2.3}
\left(
 \begin{array}{cc}
    - i \displaystyle{\frac{\Gamma_L + \Gamma_R}{2}} & \displaystyle{- \frac{\Gamma_S}{2}} \\
    - \displaystyle{\frac{\Gamma_S}{2}} & - i \displaystyle{\frac{\Gamma_L + \Gamma_R}{2}} \\
 \end{array}
\right) \; .
\end{equation}

As concerns the second contribution $\hat{\Sigma}_U^r$ we shall calculate it in the Hubbard I approximation, which should be qualitatively reliable outside the Kondo regime. Such approximation amounts to replace the matrix elements of the non-interacting Green's function (\ref{GF_T}) by
\begin{align} \label{HubbardI}
& g_{11}^r(E) = \frac{1 - \langle n_\downarrow \rangle}{E - \varepsilon_0 + i 0^+} + \frac{\langle n_\downarrow \rangle}{E - \varepsilon_0 - U + i 0^+} \; , \nonumber \\
& g_{22}^r(E) = \frac{1 - \langle n_\uparrow \rangle}{E + \varepsilon_0 + i 0^+} + \frac{\langle n_\uparrow \rangle}{E + \varepsilon_0 + U + i 0^+}. \;
\end{align}
It is important to notice that the charge densities $\langle n_\uparrow \rangle = \langle n_\downarrow \rangle = n / 2$ have to be calculated self-consistently from
\begin{align}
n & = 2 \int \frac{dE}{2 \pi} [ |G_{11}^r|^2 ( \Gamma_L f_L + \Gamma_R f_R ) \nonumber \\ \label{nU1}
& + | G_{12}^r |^2 ( \Gamma_L \tilde{f}_L + \Gamma_R \tilde{f}_R ) ] \; .
\end{align}

\section{\label{app:partialG}Partial conductances in three-terminal normal structure}

As a starting point of our two-sep protocol, we briefly analyze the partial conductances for three-terminal normal structure, within the B\"{u}ttiker approach~\cite{buttiker1986}.

Let us consider the normal electrode (\textit{S}) as a probe, assuming $J_S = 0$. The charge conservation rule implies that $J_L = - J_R \equiv J_{LR}$. From \eqref{eq:JLin2} one can find that for isothermal situation ($T_i = T$) the potential biases $\Delta \mu_{ij}$ between $i$-th and $j$-th electrodes are
\begin{align}
& \Delta \mu_{LS} = \frac{\mathcal{L}_{RS, \mu}}{D_L} J_{LR} \; , \nonumber \\ \label{eq:mu12ijt0}
& \Delta \mu_{RS} = -\frac{\mathcal{L}_{LS, \mu}}{D_L} J_{LR} \; , \\
& \Delta \mu_{LR} = \frac{\mathcal{L}_{RS, \mu} + \mathcal{L}_{LS, \mu}}{D_L} J_{LR} = \Delta \mu_{LS} - \Delta \mu_{RS} \; , \nonumber
\end{align}
where $D_L = \mathcal{L}_{LR, \mu} \mathcal{L}_{RS, \mu} + \mathcal{L}_{LR, \mu} \mathcal{L}_{LS, \mu} + \mathcal{L}_{RS, \mu} \mathcal{L}_{LS, \mu}$. The local and non-local resistances are defined as
\begin{equation}
R_{LR, ij} \equiv \frac{\Delta \mu_{ij}}{e J_{LR}}
\end{equation}
In particular, $R_{LR, LR}$ is the local resistance which can be affected by probe. The non-local resistances $R_{LR, RS} = - R_{LR, SR}$, $R_{LR, SL} = -R_{LR, LS}$ refer to the voltage between $\{ R, S \}$ or $\{ S, L \}$ terminals and the current flowing between $\{ L, R \}$ electrodes. Since energy of the system is conserved, $\Delta \mu_{LR} + \Delta \mu_{RS} + \Delta \mu_{SL} = 0$, the local resistance $R_{LR, LR}$ is a linear combination of the two non-local resistances $R_{LR, SL}$ and $R_{LR, RS}$, $R_{LR, LR} = - R_{LR, SL} - R_{LR, RS}$. The local and nonlocal conductances are equivalent to the circuit conductances consisting of resistors $1 / (e \mathcal{L}_{ij, \mu} )$ in a triangular geometry. In this way the local conductance is a sum of the direct transmission between two (\textit{L} and \textit{R}) terminals combined with an indirect one \textit{via} the additional voltage probe \textit{S}:
\begin{equation}
\mathcal{G}_{LR, LR} \equiv \frac{1}{R_{LR, LR}} = e \mathcal{L}_{LR, \mu} + \frac{e}{\cfrac{1}{\mathcal{L}_{RS, \mu}} + \cfrac{1}{\mathcal{L}_{LS, \mu}}} \; .
\end{equation}
We can notice that transmission in this additional channel is a sum of transmissions through two barriers coupled in series. In a similar way we can express the non-local conductances:
\begin{align}
\mathcal{G}_{LR, RS} & \equiv \frac{1}{R_{LR, RS}} \nonumber \\
& = - e \left( \mathcal{L}_{LR, \mu} + \mathcal{L}_{RS, \mu} + \frac{\mathcal{L}_{LR, \mu} \mathcal{L}_{RS, \mu}}{\mathcal{L}_{LS, \mu}} \right) \; , \\
\mathcal{G}_{LR, LS} & \equiv \frac{1}{R_{LR, LS}} \nonumber \\
& = e \left( \mathcal{L}_{LR, \mu} + \mathcal{L}_{LS, \mu} + \frac{\mathcal{L}_{LR, \mu} \mathcal{L}_{LS, \mu}}{\mathcal{L}_{RS, \mu}} \right) \; .
\end{align}
For experimental evaluation of the linear coefficients $\mathcal{L}_{LR, \mu}, \mathcal{L}_{RS, \mu}, \mathcal{L}_{LS, \mu}$ one has to measure three independent non-local resistances, \textit{e.g.} $R_{LR, RS}, R_{LR, LS}, R_{RS, LS}$. Besides two independent Eqs.~\eqref{eq:mu12ijt0} one needs to incorporate the third one \textit{e.g.} treating \textit{L} electrode as a probe. After some algebra one obtains
\begin{align} \label{eq:L-normal-exp}
\mathcal{L}_{LS, \mu} & = -\frac{R_{LR, RS}}{e D_R} \; , \nonumber \\
\mathcal{L}_{RS, \mu} & = \frac{R_{LR, LS}}{e D_R} \; , \\
\mathcal{L}_{LR, \mu} & = \frac{R_{RS, LS}}{e D_R} \; , \nonumber
\end{align}
with the denominator $D_R = R_{LR, LS} R_{RS, LS} - R_{LR, RS} R_{LR, LS} - R_{LR, RS} R_{RS, LS}$. Summarizing, let us notice that from~\eqref{eq:L-normal-exp} one can determine the linear coefficients $L_{ij, \mu}$ by measuring three resistances.

\section{\label{app:res}Local and non-local resistances in hybrid three-terminal structure with QD}

Consider first the scenario with the metallic electrode (say \textit{L}) acting as a voltage probe under isothermal conditions $T_L = T_R = T_S$. This means that $J_L = 0$. In the linear response regime, assuming $J_R = -J_S \equiv J_{RS}$, the local and non-local resistances are defined as
\begin{align} \label{eq:s1}
& R_{RS, RL} \equiv \frac{\Delta \mu_{RL}}{e J_{RS}} = \frac{\mathcal{L}_{LL}^{DAR} + \mathcal{L}_{LR}^{CAR}}{e D} = R_{RL, RS} \; , \\ \label{eq:s2}
& R_{RS, LS} \equiv \frac{\Delta \mu_{LS}}{e J_{RS}} = \frac{\mathcal{L}_{LR}^{ET} - \mathcal{L}_{LR}^{CAR}}{2 e D} = R_{LS, RS} \; , \\ \label{eq:s3}
& R_{RS, RS} \equiv \frac{\Delta \mu_{RS}}{e J_{RS}} = \frac{\mathcal{L}_{LR}^{ET} + 2 \mathcal{L}_{LL}^{DAR} + \mathcal{L}_{LR}^{CAR}}{2 e D} \\
& = R_{RS, RL} + R_{RS, LS} = R_{RL, RS} + R_{LS, RS} \; . \nonumber
\end{align}
Assuming the superconducting electrode to be floating (\textit{i.e.} $J_S = 0$) and denoting $J_R = -J_L \equiv J_{RL}$ we get
\begin{align} \label{eq:s4}
& R_{RL, LS} \equiv \frac{\Delta \mu_{LS}}{e J_{RL}} = - \frac{\mathcal{L}_{RR}^{DAR} + \mathcal{L}_{LR}^{CAR}}{e D} = R_{LS, RL} \; , \\ \label{eq:s5}
& R_{RL, RS} \equiv \frac{\Delta \mu_{RS}}{e J_{RL}} = \frac{\mathcal{L}_{LL}^{DAR} + \mathcal{L}_{LR}^{CAR}}{e D} = R_{RS, RL} \; , \\
& R_{RL, RL} \equiv \frac{\Delta \mu_{RL}}{e J_{RL}} = \frac{\mathcal{L}_{LL}^{DAR} + 2 \mathcal{L}_{LR}^{CAR} + \mathcal{L}_{RR}^{DAR}}{e D} \nonumber \\ \label{eq:s6}
& = R_{RL, RS} - R_{RL, LS} = R_{RS, RL} - R_{LS, RL} \; ,
\end{align}
with the same denominator
\begin{align}
D & = \mathcal{L}_{LR}^{ET} ( \mathcal{L}_{LL}^{DAR} + 2 \mathcal{L}_{LR}^{CAR} + \mathcal{L}_{RR}^{DAR} ) \nonumber \\
& + \mathcal{L}_{LR}^{CAR} ( \mathcal{L}_{LL}^{DAR} + \mathcal{L}_{RR}^{DAR}) + 2 \mathcal{L}_{LL}^{DAR} \mathcal{L}_{RR}^{DAR} \; .
\label{denominator}
\end{align}
for both cases.

\section{\label{app:currhyb}Hybrid system with a floating electrode under non-isothermal conditions}

For the floating normal \textit{L} electrode we obtain from \eqref{eq:jsc}
\begin{align} \label{eq:jscRS}
J_{RS} & = \frac{\Delta \mu_{LS}}{e R_{RS, LS}} + \frac{R_{LS, RL}}{R_{RS, LS}} ( \mathcal{L}_{LR, T}^{ET} + \mathcal{L}_{LR, T}^{CAR} ) \Delta T_{RL} \; , \\
J_{RS} & = \frac{\Delta \mu_{RS}}{e R_{RS, RS}} + \frac{R_{RS, RL}}{R_{RS, RS}} ( \mathcal{L}_{LR, T}^{ET} + \mathcal{L}_{LR, T}^{CAR} ) \Delta T_{RL} \; , \\
J_{RS} & = \frac{\Delta \mu_{RL}}{e R_{RS, RL}} + \frac{R_{RL, RL}}{R_{RS, RL}} ( \mathcal{L}_{LR, T}^{ET} + \mathcal{L}_{LR, T}^{CAR} ) \Delta T_{RL} \; .
\end{align}
In these expressions the local and non-local resistances $R_{lk, ij}$ have been determined by four probe measurements under isothermal conditions (\textit{i.e.} $T_L = T_R = T_S$). Similar equations can be derived for the floating \textit{S} electrode
\begin{align}\label{eq:jscRL}
J_{RL} & = \frac{\Delta \mu_{LS}}{e R_{RL, LS}} + ( \mathcal{L}_{LR, T}^{ET} + \mathcal{L}_{LR, T}^{CAR} ) \Delta T_{RL} \; , \\
J_{RL} & = \frac{\Delta \mu_{RS}}{e R_{RL, RS}} + ( \mathcal{L}_{LR, T}^{ET} + \mathcal{L}_{LR, T}^{CAR} ) \Delta T_{RL} \; , \\
J_{RL} & = \frac{\Delta \mu_{RL}}{e R_{RL, RL}} + ( \mathcal{L}_{LR, T}^{ET} + \mathcal{L}_{LR, T}^{CAR} ) \Delta T_{RL} \; .
\end{align}
Definitions of the resistances in both these cases are explicitly given in the Appendix~\ref{app:res} and the procedure for determination of the coefficients $L_{LR}^{ET (CAR)}$, $L_{LL (RR)}^{DAR}$ has been described by us in Ref.~[\onlinecite{michalek2015}]. In both cases the three different voltages may be measured and this hints at a possibility to define the three different thermopowers as discussed in the main text.

\cleardoublepage

\end{document}